\documentclass[a4paper,twocolumn]{article}
\usepackage{epsfig}
\usepackage{amsmath}
\usepackage{psfrag}

\title{\vspace{+0.5em} \bf{Pentaquark Masses and Magnetic Moments in a Quark Cluster Approach\\} ~\hspace{16cm}\raisebox{0cm}[0cm][0cm]{\raisebox{5.5cm}{\large}}\vspace{-1.5em}}
\author{P. Jim\'enez Delgado \\ \normalsize \textit{Department of Physics and Astronomy, Vrije Universiteit Amsterdam}\\\normalsize \textit{De Boelelaan 1081, 1081 HV Amsterdam, The Netherlands}\\[1em]
\parbox[t]{0.91\textwidth}{\small{We study the stability of the different quark substructures in a quark cluster approach to pentaquark states considering the color-magnetic spin-spin interactions between quarks. The most likely configu- ration is found to be a triquark-diquark one where the two quarks of the diquark are coupled to spin zero and anti-triplet representations of flavor and color, and the triquark has spin $\frac{1}{2}$ and belongs to the triplet representation of color and to anti-sextet of flavor. To ensure total antisymmetry under the interchange of identical particles, triquark and diquark are in relative p-wave. Using this configu- ration we estimate the masses and magnetic moments of pentaquarks. Finally the calculation of the masses has been extended to some charmed pentaquarks.\vspace{-1em}}}}
\date{}

\begin{document}
\fontsize{9}{10.7}\selectfont
\flushbottom
\maketitle

\section*{Introduction}\label{introduction}
In recent papers [1~-~8], it has been reported the detection of a new manifestly exotic baryon resonance with the quantum numbers of a nucleon plus a kaon (having charge $Q=+1$, baryon number $B=+1$, isospin $I_3=0$ and strangeness $S=+1$). The width of the state was limited in all the observations by the experimental resolution, but it has been estimated to be very narrow (4 - 25 MeV).
Although it is possible that such a resonance may be regarded as a meson-baryon state ($pK^0$ or $nK^+$) bound by molecular-like color forces, the narrow width suggest a complex internal structure. In particular an exotic baryon with these quantum numbers is also a natural candidate.

The structure of hadrons should be studied properly within the \emph{Standard Model} using the theory for the strong interaction (\emph{QCD}). However, this theory is still not well understood at low energy regimes (bound states), where it is not perturbative. Consequently simpler phenomenological approaches like \emph{constituent quark} models, \emph{bag} models and \emph{Chiral-Soliton} models have been used to study these states.

\begin{figure}[h]
\begin{center}
\begin{minipage}{0.45\textwidth}
\begin{center}
    \psfrag{1}[cc][cc]{\footnotesize $uudd\overline{s}$~$(\Theta^+)$}
    \psfrag{2}[cc][cc]{\footnotesize$udds\overline{s}$}
    \psfrag{3}[cc][cc]{\footnotesize$uudd\overline{u}$}
    \psfrag{4}[cc][cc]{\footnotesize$uudd\overline{d}$}
    \psfrag{5}[cc][cc]{\footnotesize$uuds\overline{s}$}
    \psfrag{6}[cc][cc]{\footnotesize$udds\overline{u}$}
    \psfrag{7}[cc][cc]{\footnotesize$ddss\overline{s}$}
    \psfrag{8}[cc][cc]{\footnotesize$uuds\overline{u}$}
    \psfrag{9}[cc][cc]{\footnotesize$udss\overline{s}$}
    \psfrag{10}[cc][cc]{\footnotesize$udds\overline{d}$}
    \psfrag{11}[cc][cc]{\footnotesize$uuds\overline{d}$}
    \psfrag{12}[cc][cc]{\footnotesize$uuss\overline{s}$}
    \psfrag{13}[cc][cc]{\footnotesize$uuss\overline{d}$~$(\Xi^+)$}
    \psfrag{14}[cc][cc]{\footnotesize$uuss\overline{u}$}
    \psfrag{15}[cc][cc]{\footnotesize$udss\overline{d}$}
    \psfrag{16}[cc][cc]{\footnotesize$udss\overline{u}$}
    \psfrag{17}[cc][cc]{\footnotesize$ddss\overline{d}$}
    \psfrag{18}[cc][cc]{\footnotesize$ddss\overline{u}$~$(\Xi^{--})$}
    \psfrag{a}[cc][cc]{\small$Y$}
    \psfrag{b}[cc][cc]{\small$I_3$}
    \psfrag{c}[cc][cc]{\small$Q=-2$}
    \psfrag{d}[cc][cc]{\small$Q=-1$}
    \psfrag{e}[cc][cc]{\small$Q=0$}
    \psfrag{f}[cc][cc]{\small$Q=1$}
\epsfig{file=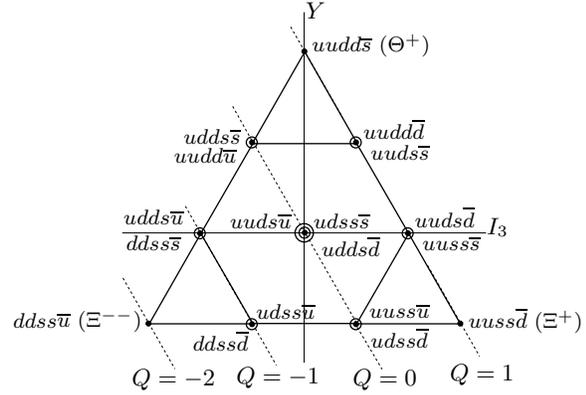,width=0.85\textwidth}
\caption{\small $SU(3)_F$ representations of anti-decuplet and octets for four quarks and an antiquark (the triangle side is 3 units long).\vspace{-1em}}
\label{antidecuplet}
\end{center}
\end{minipage}
\end{center}
\end{figure}

Using the \emph{Chiral-Soliton} model Diakonov \emph{et al.} \cite{Diakonov:1997mm} predicted in 1997 the existence of an exotic anti-decuplet of baryons (Fig. \ref{antidecuplet}) containing exotic states (the corners of the triangle). This prediction stimulated the active search of the reported resonance and its final discovery. The quark content of this hypothetical exotic baryon was proposed to be $uudd\overline{s}$ so that it is called a pentaquark and has been assigned the symbol $\Theta^+$. Recently a second exotic pentaquark resonance $\Xi^{--}$  with minimal quark content $ddss\overline{u}$ has been reported  \cite{Alt:2003vb}. There is also a possible observation of a charmed pentaquark $\Theta^0_c$ in an old experimental \mbox{paper~\cite{Amirzadeh:1979qi}}.

Although the prediction of the $\Theta^+$ was made in the framework of the \emph{Chiral-Soliton} Model, there is an interesting active discussion about the configuration of pentaquarks in an effective theory of constituent quarks. Even within this naive picture there exist several possible configurations for the structure of pentaquarks depending on the quark substructures that are considered. The obvious one is to consider them as a single cluster of uncorrelated quarks. Although this option is not so popular nowadays because it fails to explain why the lifetime of the states is so large, it has been used in \emph{bag model} calculations \cite{Strottman:1979qu,deSwart:1981fn}. Other quark cluster configurations that have been proposed consist in two strongly correlated diquarks (clusters made up of two quarks) bound together with the antiquark \cite{Jaffe:2003sg, Shuryak:2003zi} or in a diquark bound with a triquark (two quarks and an antiquark) \cite{Karliner:2003dt}. A different triquark-diquark model for the structure of pentaquarks is proposed in this article as consequence of the stability of quark clusters under the short-range color-magnetic spin-spin interactions.
\vspace{-1em}

\section*{Quark Cluster Approach}\label{cluster}
The simplest approach to the study of hadrons is an effective description in terms of massive ($\sim m_p$) constituent quarks which may form quark substructures (clusters). 

The quantum numbers of an hadron are those of the combinations of the so-called \emph{valence} quarks. Due to the nature of \emph{Quantum Field Theory}, as a consequence of the strong coupling many pairs of particle-antiparticle may be created which are known as \emph{sea} quarks. Thus the true vacuum (lowest energy) state within a hadron will be a complicated state of gluons and valence and sea quarks and antiquarks. It can be argued \cite{Gasiorowicz:1981jz} that within this medium, the long range color interaction between valence quarks is mainly independent of quark spins and quark masses, depending only on the spatial separation of these quarks and their colors.

We assume that the principal binding forces of hadrons are these long-range confining forces so that the confinement mechanism together with the relativistic dynamics of quarks within a hadron, make the masses of valence quarks appear like \emph{constituent quark masses} which may be obtained by means of experimental information and therefore are model dependent. There is an uncertainty of about a hundred MeV in the estimation of these constituent quark masses (the masses that we use are given in Table~\ref{cmasses}) and therefore in the calculations realized using this naive approach.

Internal excitations (rotations, etc) may exist within hadrons. In this article, unless it is necessary for fundamental reasons (which actually is the case) we will deal only with what are called \emph{ground state} hadrons, i.e, hadrons for which the state of any internal degree of freedom is of minimal energy (for instance in the case of rotations all quarks would be in an \mbox{s-wave} if it is possible, otherwise in a \mbox{p-wave}, etc).

Once the \emph{long-range} confining part of the strong interaction has been taken into account, further refinements in the description of hadrons may be considered by means of \emph{short-range} interactions between \emph{valence}  quarks. In particular we will consider the color-magnetic interactions analogous to the electromagnetic spin-orbit and spin-spin interactions which produces respectively the fine and the hyperfine structure in the atomic spectrum. In the hadronic case \cite{DeRujula:1975ge}, the spin-spin color interaction turns out to be relatively more important (some hundreds MeV) than the spin-orbit color interaction (generally less than a hundred MeV). Considering a fixed number of quarks and antiquarks forming a color singlet, the order of the states in the spectrum is indicated by the spin-spin interaction between quarks. Besides, for pentaquarks the magnitude of the spin-orbit color interaction is expected to be smaller than the uncertainty in constituent masses (about a hundred MeV) and therefore we will not consider it.

The origin of the hyperfine structure is the color-magnetic field generated by the quarks, which depends in general on their separation, masses, spin and color. For multi-quark systems the color spin-spin interaction energy of the system from lowest-order one gluon exchange is \cite{Jaffe:1976ih, Mulders:1979ea}:
\begin{equation}
\Delta E_{ss} = - \sum_{a=1}^{8}\sum_{i>j} M(m_i,m_j)~ (F_i^aF_j^a)(\vec{\sigma_i}\cdot\vec{\sigma_j})
\label{finemul}
\end{equation}
where $\vec{\sigma_i}$ are the spin matrices and $F_i^a$ are the color fields of the \emph{i}th constituent quark (which are summed over the group index $a$) and $M(m_i,m_j)$ measures the interaction strength.

The exact evaluation of the hyperfine splitting is quite complicated and usually some approximations are made (factorizing, averaging over possible configurations, etc. See \cite{Jaffe:1976ih, Mulders:1979ea} for a more detailed treatment). We will use an approximate form for the hyperfine splitting from bag model calculations from \cite{Mulders:1979ea}:
\begin{equation}
\begin{split}
&\Delta E_{ss}=m(N,N_s)\Delta \hspace{0.5cm} \mbox{with} \\
&\Delta = -\sum_{i>j}\sum_a(F_i^aF_j^a)(\vec{\sigma_i}\cdot\vec{\sigma_j})
\label{fine}
\end{split}
\end{equation}

The strength of the color magnetic interactions $m(N,N_S)$ depends in this approximation on the number of quark in the system $N$ and how many of them are strange $N_s$ (although for a fix $N$ its values differ only by a few percent for different $N_S$, see Table~\ref{strength}) . Aside from this, the strength is independent of the configuration, being collected all this dependence of the interaction in the so-called group theoretical factor $\Delta$, which may be calculated exactly from the spin and color representations of the quark clusters. Based on the similarity of the different values for the strength, we may discuss the importance of this interaction in terms of the group theoretical factor $\Delta$ instead of the actual interaction energy $\Delta E_{ss}$.

The physical states of a cluster are identified with the eigenvectors of its mass operator, therefore, under the approximation of (\ref{fine}), they are the eigenvectors of the group theoretical factor. $\Delta$ depends in general on the $SU(6)_{CS}\subset SU(2)_S\otimes SU(3)_C$ irreducible representations of the set of quarks ($Q$) and antiquarks ($\overline{Q}$) separately and of the entire system ($Q\overline{Q}$). We will note these representations by $f_{CS}\equiv[f_1,f_2,...f_5]$ (where $f_i$ is the number of blocks in the row $i$ corresponding to the Young tableau of the representation), $S$ and $d_C$ respectively (where $d$ indicates the dimension of the representation). The letter $f$ is used for the color-spin representation because the $SU(6)_{CS}$ color-spin representation of a set of identical particles has to be the conjugate of its $SU(3)_{F}$ flavor representation to ensure antisymmetry (assuming that they are in relative s-wave) of the total wave function on identical particles, therefore it depends on the flavor degrees of freedom. This does not determine completely the $SU(6)_{CS}$ representation for the total system, the total state is necessarily a color singlet and has a definite spin, but does not in general belong to a $SU(6)_{CS}$ representation.

To describe the total state two basis may be used. A natural basis convenient for the calculation of the magnetic moments is the tensorial product of states describing the quark and antiquark clusters $|\Phi_l\rangle\equiv|Q:f_Q,d_Q,S_Q\rangle\otimes|\overline{Q}:f_{\overline{Q}},d_{\overline{Q}},S_{\overline{Q}}\rangle$, which diagonalize the $C_2$, $C_3$ and $C_6$ quadratic Casimir operators of quarks and antiquarks separately as well as the $C_2$ and $C_3$ operators of the entire system. Another basis is made up of the states which diagonalize the $C_6$ (and $C_2$  and $C_3$) quadratic Casimir operator of the entire system, i.e, states belonging to a particular $SU(6)_{CS}$ representation $|\Psi_i\rangle\equiv|Q\overline{Q}:f_{Q\overline{Q}},d_{Q\overline{Q}},S_{Q\overline{Q}}\rangle$. Their are related by the $SU(6)_{CS}\subset SU(2)_S\otimes SU(3)_C$ \emph{Wigner} (\emph{Clebsh-Gordan}) coefficients ($M_{il}\equiv \langle\Phi_l|\Psi_i\rangle$) which may be calculated using the method from \cite{Strottman:1979et}.

An expression for the matrix element of the group theoretical factor of a cluster of $N$ particles has been worked out in \cite{Strottman:1979qu}:

\begin{equation}
\begin{split}
&(\Delta_\Psi)_{ij}=\langle\Psi_i|\Delta|\Psi_j\rangle=\\ &\hspace{3em} =\sum_l\langle\Phi_l;\Psi_i|\Delta|\Phi_l;\Psi_j\rangle\langle\Phi_l|\Psi_i\rangle\langle\Phi_l|\Psi_j\rangle~\mbox{ where }\\
&\langle\Phi_l;\Psi_i|\Delta|\Phi_l;\Psi_j\rangle=[2N+\frac{1}{2}C_{6Q\overline{Q}}(\Psi_i)-\frac{1}{2}C_{3Q\overline{Q}}(\Psi_i)\\
&\hspace{6em}-\frac{1}{3}C_{2Q\overline{Q}}(\Psi_i)-C_{6Q}(\Phi_l)-C_{6\overline{Q}}(\Phi_l)]\delta_{ij}\\
&\hspace{3em}+C_{3Q}(\Phi_l)+C_{3\overline{Q}}(\Phi_l)+\frac{2}{3}C_{2Q}(\Phi_l)+\frac{2}{3}C_{2\overline{Q}}(\Phi_l)
\end{split}
\end{equation}
Once we have the operator expressed in this basis, we can change to the basis of the product states by using the $SU(6)_{CS}\subset SU(2)_S\otimes SU(3)_C$ CG coefficients (\mbox{$\vec{\Psi}=M\vec{\Phi}$}, \mbox{$\Delta_{\Phi}=M^{-1}\Delta_{\Psi}M$}). Upon diagonalization of the operator, the eigenvalues are the values for $\Delta$ and the eigenvectors the physical states expressed in the basis of the product of states for quarks and antiquarks separately.

Some product states $|\Phi_l\rangle$ can only be coupled to a total state $|\Psi_i\rangle$,i.e, \mbox{$M_{il}\equiv \langle\Phi_l|\Psi_i\rangle=1$} and the system belong to a unique $SU(6)_{CS}$ representation. The group theoretical factor  for these states is already diagonal in both (identical) bases and its expression reduces to:
\begin{equation}
\begin{split}
\Delta\equiv(\Delta_{\Psi})_{ii}=&2N-\frac{1}{2}\Delta_{Q\overline{Q}}+\Delta_{Q}+\Delta_{\overline{Q}}~\mbox{ with }\\
&~\Delta_X\equiv C_{3X}+\frac{2}{3}C_{2X}-C_{6X}
\label{deltaj}
\end{split}
\end{equation}
Which is the expression given in \cite{Jaffe:1976ih} (up to a factor of 4) for the group theoretical factor of a s-wave hadronic system. Expressions for the Casimir operators may be found in \cite{Strottman:1979qu, Jaffe:1976ih} as well. The group theoretical factor for many multi-quark systems has already been calculated (see for instance \cite{Mulders:1979ea, Aerts:1979hn}).
\vspace{-1em}

\section*{Constituent Quark Structures} \label{structures}
A pentaquark configuration may be realized by different quark substructures as it has been proposed in \cite{Jaffe:2003sg,Shuryak:2003zi, Karliner:2003dt}. In order to study the stability of these structures, we consider that the distances between valence (constituent) quarks may change dynamically by short distances, so that the short-range interactions (namely, the color-magnetic spin-spin interaction) may change considerably without affecting notably the long-range confining forces. For different configurations, the color-magnetic interaction is in general different, so that the most stable of these structures are the ones with lowest color-magnetic energy (or equivalently lowest $\Delta$) and the ones that are expected to be present within hadrons.

Other authors \cite{Cheung:2003de} consider further hyperfine color-magnetic splitting between the clusters building up pentaquarks. We consider the color-magnetic interactions to be short-range forces so that constituent quarks in different clusters do not interact through this force. According to this, the color-magnetic interaction is not related to the confinement mechanism, which is due to the color-electric force and is completely reflected in the constituent quark masses.

To obtain $\Delta$ we have to calculate the different group theoretical configurations for the possible structures. Let us consider first a single cluster of uncorrelated quarks. The color representation for a pentaquark needs to be a singlet and the representation for the antiquark is $\overline{3}_C$ (\parbox{0.2cm}{\epsfig{file=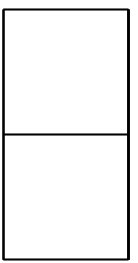,width=0.2cm}}), therefore the color representation for the four quarks needs to be $3_C$ (\parbox{0.4cm}{\epsfig{file=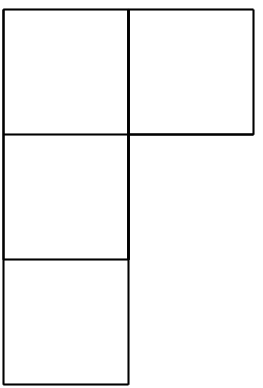,width=0.4cm}}). Since the four quarks are identical fermions we have to apply Pauli's exclusion principle to the complete state, this implies that the four quarks need to belong to the $210_{FS}$ representation of the symmetry group $SU(6)_{FS}\supset SU(3)_F\otimes SU(2)_S$.

As we have seen, pentaquarks are assumed to belong to the anti-decuplet flavor representation $\overline{10}_F$ (\parbox{0.6cm}{\epsfig{file=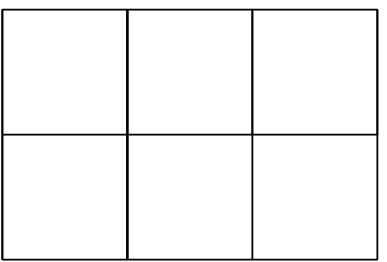,width=0.6cm}}). This representation comes from the $SU(3)_F$ product of representations $\overline{6}\otimes\overline{3}$, where $\overline{6}_F$ (\parbox{0.4cm}{\epsfig{file=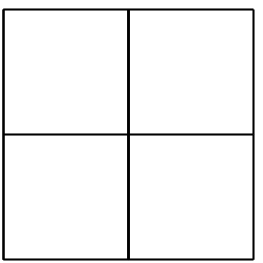,width=0.4cm}}) is the flavor representation for the four quark state. Therefore, to obtain a $210_{FS}$ representation we need a $3_S$ (\parbox{0.6cm}{\epsfig{file=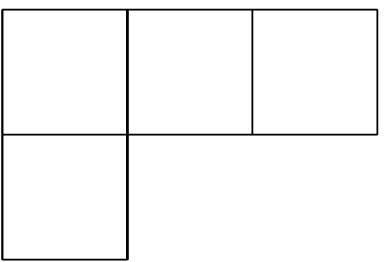,width=0.6cm}}) representation of  the spin $SU(2)_S$ group of the four quarks, i.e. the four quarks couple to spin $1$. We can then write the configuration of the four quarks belonging to the pentaquark in the form:
\begin{equation}
|Q^4\rangle= \parbox{0.4cm}{\epsfig{file=./pictures/6b.eps,width=0.4cm}}
\otimes \parbox{0.6cm}{\epsfig{file=./pictures/s3p4.eps,width=0.6cm}}
\otimes \parbox{0.4cm}{\epsfig{file=./pictures/3p4.eps,width=0.4cm}}
=\overline{6}_F\otimes3_S\otimes3_C
\label{q4for}\end{equation}

We already known that the flavor representation for a pentaquark is $\overline{10}_F$ and the color representation is $\overline{1}_C$. The spin of the antiquark couples to the spin of the four quark either into a spin representation $2_S$ (spin $\frac{1}{2}$) or  into a $4_S$ representation (spin $\frac{3}{2}$). Therefore, we have two different kinds of wave functions for uncorrelated quarks, one describing the $\Theta^+$, $\Xi^{--}$ and $\Xi^+$ pentaquarks and another describing the $\Theta^{*+}$, $\Xi^{*--}$ and $\Xi^{*+}$. These
configurations and their group theoretical factors are respectively:
\begin{equation}
\hspace{-1em}
\begin{array}{l}
|Q^4\overline{Q}\rangle_1=\parbox{0.6cm}{\epsfig{file=./pictures/10b.eps,width=0.6cm}}
\otimes\parbox{0.6cm}{\epsfig{file=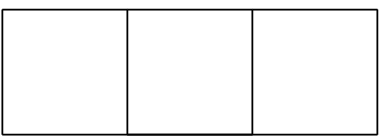,width=0.6cm}}
\otimes\parbox{0.4cm}{\epsfig{file=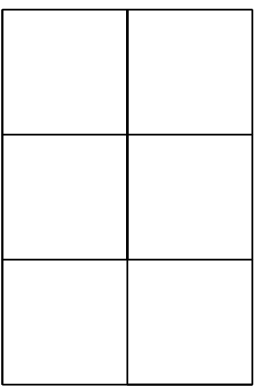,width=0.4cm}}
=\overline{10}_F\otimes2_S\otimes1_C\rightarrow \Delta=-\frac{14}{3}
\\[0.4cm]
|Q^4\overline{Q}\rangle_2=\parbox{0.6cm}{\epsfig{file=./pictures/10b.eps,width=0.6cm}}
\otimes\parbox{0.2cm}{\epsfig{file=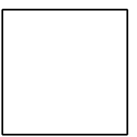,width=0.2cm}}
\otimes\parbox{0.4cm}{\epsfig{file=./pictures/1p5.eps,width=0.4cm}}
=\overline{10}_F \otimes 4_S\otimes 1_C\rightarrow \Delta=+\frac{1}{3}
\end{array}
\label{uncorrfor}
\end{equation}

A pentaquark state may also be realized by bound states of quark clusters. Let us consider first diquark ($qq$) structures. According to Pauli's exclusion principle two quarks have to be in an antisymmetric total wave function, this leads to the following group theoretical representations and group theoretical factors:
\begin{equation}
\begin{array}{l}
|Q^2\rangle_1=\overline{3}_F\otimes1_S\otimes\overline{3}_C\rightarrow \Delta=-2\\[0.1cm]
|Q^2\rangle_2=\overline{3}_F\otimes3_S\otimes6_C\rightarrow \Delta=-\frac{1}{3}\\[0.1cm]
|Q^2\rangle_3=6_F\otimes3_S\otimes\overline{3}_C\rightarrow \Delta=+\frac{2}{3}\\[0.1cm]
|Q^2\rangle_4=6_F\otimes1_S\otimes6_C\rightarrow \Delta=+1
\end{array}
\label{diquarks}
\end{equation}
Therefore, if one of the substructures within a pentaquark is a diquark, we expect that it will take the configuration of lowest group theoretical factor that is possible, i.e., the first one in (\ref{diquarks}), otherwise the second one and so on. For three quarks (an ordinary baryon) the configurations and the group theoretical factor are known to be:

\begin{equation}
\begin{array}{l}
|Q^3\rangle_1=
\parbox{0.4cm}{\epsfig{file=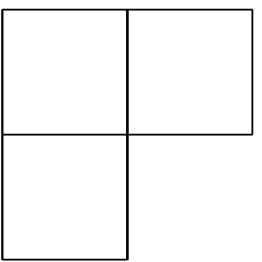, width=0.4cm}}\otimes
\parbox{0.4cm}{\epsfig{file=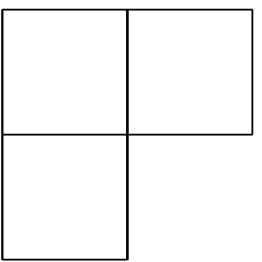, width=0.4cm}}\otimes
\parbox{0.2cm}{\epsfig{file=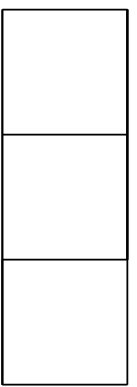, width=0.2cm}}
=8_F\otimes2_S\otimes1_C\rightarrow \Delta=+2
\\[0.2cm]
|Q^3\rangle_2=
\parbox{0.6cm}{\epsfig{file=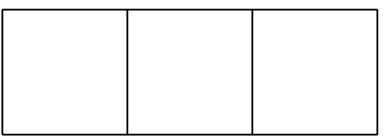, width=0.6cm}}\otimes
\parbox{0.6cm}{\epsfig{file=./pictures/s4.eps, width=0.6cm}}\otimes
\parbox{0.2cm}{\epsfig{file=./pictures/1.eps, width=0.2cm}}=10_F\otimes4_S\otimes1_C\rightarrow \Delta=-2
\end{array}\label{baryon}\end{equation}

For four quarks of a pentaquark $\Delta=-\frac{4}{3}$ so that the formation of two diquarks is energetically favorable, thus we do not expect four-quark clusters within pentaquarks.

The rest of clusters that may be formed with the quark content of a pentaquark include the antiquark. For instance the configurations and group factors of  a quark and an antiquark (mesons) are:
\begin{equation}
\begin{array}{l}
|Q\overline{Q}\rangle_1=8_F\otimes1_S\otimes1_C\rightarrow \Delta=-\frac{12}{3}\\[0.1cm]
|Q\overline{Q}\rangle_2=8_F\otimes3_S\otimes1_C\rightarrow \Delta=+\frac{4}{3}
\end{array}
\label{formeson}
\end{equation}
If pentaquarks are made up of a diquark and a triquark ($qq\overline{q}$) we consider the diquark to be in a $\overline{3}_F$ of flavor, the triquark therefore needs to be in a  $\overline{6}_F$ so that they can form the $\overline{10}_F$ of a pentaquark. This leads to the following triquark configurations:
\begin{equation}
\begin{array}{l}
|Q^2\overline{Q}\rangle_1=\overline{6}_F\otimes2_S\otimes3_C\rightarrow \Delta=-\frac{16.26}{3}\\[0.1cm]
|Q^2\overline{Q}\rangle_2=\overline{6}_F\otimes2_S\otimes\overline{6}_C\rightarrow  \Delta=-\frac{6}{3}\\[0.1cm]
|Q^2\overline{Q}\rangle_3=\overline{6}_F\otimes2_S\otimes3_C\rightarrow \Delta=-\frac{0.75}{3}\\[0.1cm]
|Q^2\overline{Q}\rangle_4=\overline{6}_F\otimes4_S\otimes3_C\rightarrow  \Delta=-\frac{4}{3}
\end{array}
\label{triquarks}
\end{equation}
The states $1$ and $3$ belongs to the same final representations in all spaces but their internal structure is different. In particular they are eigenvectors of $\Delta$ made up of different linear combinations of the same product states. This will be cleared up in the next section, where we analize the triquark structure.

We can proceed in a similar way for the case of a tetraquark ($qqq\overline{q}$) obtaining as most likely configurations:
\begin{equation}
\begin{array}{l}
|Q^3\overline{Q}\rangle_1=15_F\otimes1_S\otimes\overline{3}_C\rightarrow \Delta=-\frac{12}{3}\\[0.1cm]
|Q^3\overline{Q}\rangle_2=15_F\otimes3_S\otimes\overline{3}_C\rightarrow \Delta=-\frac{14}{3}
\end{array}
\label{tetraquarks}
\end{equation}
For four quarks and an antiquark (pentaquark) the configurations and group theoretical factors have already been given in (\ref{uncorrfor}). For all these clusters there exist in general more possible configurations within a pentaquark, but all of them corresponds to higher $\Delta$.

Once we have the group theoretical factors for the possible constituent clusters, we add up the lowest among these quantities in order to roughly compare the stability of the possible arrangements; the numbers are given in Table~\ref{deltas}.
\begin{table}[h]
\begin{center}
\begin{minipage}{0.45\textwidth}
\begin{center}
\begin{tabular}[t]{|c|c|}\hline
Configuration&$\Delta$\\ \hline\hline
$\rule[-7pt]{0pt}{20pt}${\small$Q^3-Q\overline{Q}$}&$-\frac{18}{3}$\\ \hline
$\rule[-7pt]{0pt}{20pt}${\small$Q^4\overline{Q}$}&$-\frac{14}{3}$\\ \hline
$\rule[-7pt]{0pt}{20pt}${\small$Q^3\overline{Q}-Q$}&$-\frac{12}{3}$\\ \hline
$\rule[-7pt]{0pt}{20pt}${\small$Q^2\overline{Q}-Q^2$}&$-\frac{22.26}{3}$\\ \hline
$\rule[-7pt]{0pt}{20pt}${\small$Q^2-Q^2-\overline{Q}$}&$-\frac{12}{3}$\\ \hline
\end{tabular}
\caption{\small Sum of group theoretical factors $\Delta$ of the quark clusters of several possible configurations}
\label{deltas}
\vspace{-1.5em}
\end{center}
\end{minipage}
\end{center}
\end{table}

It can be seen that the most likely configuration for \mbox{pentaquarks} is the triquark-diquark option of \mbox{$\Delta=-\frac{22.26}{3}$} (i.e. $\Delta=-5.42-2$) which formal wave functions are:
\begin{equation}
|Q^2\rangle=\overline{3}_F\otimes1_S\otimes\overline{3}_C
\hspace{1cm}
|Q^2\overline{Q}\rangle=\overline{6}_F\otimes2_S\otimes3_C
\label{tridiquark}
\end{equation}

\section*{The Triquark-Diquark Model} \label{Tridiquark}
In view of the results of last section, we may propose a quark cluster model for pentaquarks consisting of a triquark-diquark model. The diquark ($\Delta=-2$) has a completely antisymmetric wave function in flavor, spin and color, so that they are quarks of different flavor ($ud$ in the case of the $\Theta^+$) coupled to spin zero and $\overline{3}$ of color.

The group theoretical factor $\Delta$ of the triquark is not diagonal in the basis of product states of two-quark clusters times the antiquark state. The eigenvector corresponding to the eigenvalue $\Delta=-5.42$ is a linear combination (\ref{triquark}) of two different ($1$ and $2$) two-quark states from (\ref{diquarks}): The one of the diquark ($\Delta=-2$) and the two-quark state of $\Delta=-\frac{1}{3}$, which quarks are coupled antisymmetrically in flavor but symmetrically in spin and color. The triquark state is the product of this linear combination and the antiquark state (see \cite{Aerts:1979hn}):
\begin{equation}
\begin{split}
|Q^2\overline{Q}\rangle=(0.582|Q^2\rangle_1-0.813|Q^2\rangle_2)\otimes|\overline{Q}\rangle\\
\mbox{~with~}\left\{\begin{split}
&~~|\overline{Q}\rangle=|\overline{3}_F,2_S,\overline{3}_C\rangle=|[1^5]_{CS},\overline{3},\frac{1}{2} _S\rangle\\
&|Q^2\rangle_1= |\overline{3}_F,1_S,\overline{3}_C\rangle= |[2]_{CS},\overline{3}_C,0_S\rangle\\
&|Q^2\rangle_2= |\overline{3}_F,3_S,6_C\rangle= |[2]_{CS},6_C,1_S\rangle\\
\end{split}\right.\\
\end{split}
\label{triquark}
\end{equation}

Both two-quark systems couple with the antiquark into a spin $S=\frac{1}{2}$ and $3_C$ color representation. Since their wave functions are completely antisymmetric under the interchange of identical particles, the wave function of the triquark is antisymmetric as well. The wave function of the diquark is also antisymmetric, so that to ensure total antisymmetry under the interchange of identical particles triquark and diquark must be in an antisymmetric spatial wave function or relative p-wave ($l=1$). Therefore, the final state is a singlet in color, belongs to the antidecuplet representation of flavor and may have spin-parity state $S^{P}=\frac{1}{2}^+$ or $S^{P}=\frac{3}{2}^+$, actually both resonances are expected in principle.

One might think that the observed exotic resonance could be a \emph{molecular} kaon-nucleon bound state. For this to be the case the hyperfine splitting of the kaon-nucleon system should be the lowest among the ones of the possible arrangements, otherwise it would be more likely for the system to take a state of lower energy without any noticeable change in the long-range binding forces. From Table~\ref{deltas}, the sum of the group theoretical factors of a kaon and a nucleon is $\Delta=-\frac{18}{3}$, which is higher than the value of the proposed triquark-diquark model. Hence, to reach the configuration of the decay ($KN$), the system has to be reconfigured against the color-magnetic interactions. This may explain the narrow width of the observed pentaquark resonances.

Arguing that the attraction between quarks in the $\overline{3}_c$, $J^P=0^+$ channel is known to be strong, \emph{Jaffe} and \emph{Wilczek} \cite{Jaffe:2003sg} proposed a model for pentaquarks based on diquarks. In particular they suggested a diquark-diquark-antiquark model where both diquarks are identical with the configuration of the pairs of quarks in the model proposed here \mbox{ ($\overline{3}_F\otimes1_S\otimes\overline{3}_C$)} and therefore it is also necessary to introduce a \mbox{p-wave} between them, so that the parity of pentaquarks in this model is plus as well. Although the strong correlation of the diquark is important for the stability of the system, it is energetically even more favorable that the antiquark sits closer to one of the diquarks forming the triquark that is proposed here.

To avoid introducing a rotational contribution (\mbox{$l\neq0$}) to the mass, it have been proposed \cite{Shuryak:2003zi} that one of the diquarks could be a tensor diquark (therefore of parity $-$), there are several possible configurations for this diquark (See eq. \ref{diquarks}), but all of them would have higher color-magnetic contributions than the diquark used by \emph{Jaffe} and \emph{Wilczek}.

A different triquark-diquark model has been proposed by Karliner and Lipkin \cite{Karliner:2003dt}. They argued that the hyperfine interaction is always repulsive between two quarks of the same flavor and propose that the diquark is in the same configuration that the used here ($|Q^2\rangle_1$) but the two quarks within the triquark are in $|Q^2\rangle_2$. In order to separate the two quark pairs they also introduce a relative \mbox{p-wave} between the clusters, so that the parity of the state is plus as well.

It is worth remarking that all these proposed models and the Chiral-Soliton model predict positive parity for the pentaquarks as opposed to the uncorrelated \mbox{s-wave} quark model and results from lattice QCD \cite{Csikor:2003ng, Sasaki:2003gi}. An experimental result on this crucial issue has not yet been obtained.
\vspace{-1em}

\section*{Masses and Magnetic Moments}
At this point we can roughly estimate pentaquark masses using this triquark-diquark model. The mass formula for a hadron in this naive quark cluster approach is:
\begin{equation}M=\sum_i m_i+\Delta E_{ss}+\Delta E_{l}\label{massformula}\end{equation}
where the sum is over \emph{valence} quarks and the masses $m_i$ are constituent quark masses, the term $\Delta E_{ss}$ represents the contribution from the spin-spin color-magnetic interactions and $\Delta E_{l}$ is the contribution to the mass due to the \mbox{p-wave}.
The constituent quark masses are obtained by fitting a set of experimental data for some hadrons and therefore are model-dependent, we will use the constituent masses which fit the low-lying (without excitations) hadron spectrum (baryon and meson spectrum respectively) (Table~\ref{cmasses}) which are given in \cite{Gasiorowicz:1981jz}.
\begin{table}[h]
\begin{center}
\begin{minipage}[t]{0.45\textwidth}
\begin{center}
\begin{tabular}[t]{|c|c|c|}\hline
Flavor& Baryon masses&Meson masses\\\hline\hline
u, d& 363 &310   \\ \hline
s & 538 & 483   \\ \hline
c & 1704 & 1662   \\ \hline
\end{tabular}
\caption{\small Constituent quark masses (in MeV) for low-lying hadrons (from \cite{Gasiorowicz:1981jz}).}
\label{cmasses}
\vspace{-2em}
\end{center}
\end{minipage}
\end{center}
\end{table}

Although it is somewhat arbitrary which masses to use, we use the masses for baryons for the triquark and the masses of mesons for the diquark in the hope that the confining color interactions are more similar between these objects. This clearly implies that there is an uncertainty of the order of hundred MeV in all the predictions made using this effective model.

The strength of the spin-spin color-magnetic interaction that we use is fitted to baryon masses using bag model results from \cite{Mulders:1979ea}, they are given in Table~\ref{strength}.
\begin{table}[h]
\begin{center}
\begin{minipage}{0.45\textwidth}
\begin{center}
\begin{tabular}[t]{|c|c|c|c|}\hline
\begin{picture}(17,10)(0,0)
\begin{scriptsize}
\put(-3,-2){$N$}
\put(23,-4.4){\line(-2,1){29}}
\put(10,3){$N_S$}
\end{scriptsize}
\end{picture}
&0&1&2\\ \hline
2&85.1&70.2&58.2\\ \hline
3&74.4&64.3&55.5\\ \hline
4&67.6&60.1&53.3\\ \hline
5&62.7&56.8&51.2\\ \hline
\end{tabular}
\caption{\small The strength $m(N,N_S)$ of the color-magnetic interaction in MeV (from \cite{Mulders:1979ea}).}
\label{strength}
\vspace{-2.5em}
\end{center}
\end{minipage}
\end{center}
\end{table}

We can obtain a rough value for the rotational energy using some experimental information from a two-body system (meson). We will use the estimate made by Karliner and Lipkin in \cite{Karliner:2003dt}. They use the recently discovered charmed resonance $D_s(2317)$ which is consider to be a \mbox{p-wave} excitation of the ground state $D_s(1969)$. After subtracting the hyperfine splitting between the states the rotational energy is $\Delta E_L=207$ MeV and the reduced mass of the system ($c\overline{s}$) is estimated to be $\mu_{c\overline{s}}=410$ MeV. Other values have been used for this energy, what clears up that there is an uncertainty of the order of a hundred MeV in these kind of estimate.

The rotational energy is the same for the two possible configurations ($S=\frac{1}{2}$ or $S=\frac{3}{2}$) so that we can not distinguish between the masses of these states. It may be possible to distinguish between them by means of the spin-orbit color interaction but this interaction has not been considered because its contribution to the masses is expected to be smaller than the uncertainty in the constituent masses and in the rotational energy (about hundred MeV).

We can calculate the masses of the constituent diquark and triquark by using eq.~(\ref{massformula}) and the strength of the color-magnetic interaction of Table~\ref{strength} together with the group theoretical factor of diquark and triquark. The results of this calculation are given in Table~\ref{ditrimass}.
\begin{table}[h]
\begin{center}
\begin{minipage}{0.45\textwidth}
\begin{center}
\begin{tabular}[t]{|c|c|c|}\hline
&Mass&$\mbox{Mass}_{\mbox{\tiny exp}}$\\ \hline\hline
$\Theta, \Theta^*$&1570&1540\\ \hline
$\Xi, \Xi^*$&1775&1860\\ \hline
\end{tabular}
\caption{\small Estimate (in MeV) of pentaquark masses using the triquark-diquark model proposed here and a rotational energy from \cite{Karliner:2003dt}.}
\label{ditrimass}
\vspace{-1.5em}
\end{center}
\end{minipage}
\end{center}
\end{table}

This numbers are different from the observed masses by a few percentage only, although the uncertainty in the results (some hundred MeV) is so large that this close agreement might well be chance.

Since in the past magnetic moments predictions from this sort of models have been well accomplished, in order to distinguish among the possible configurations, it seems to be useful to study also the magnetic moments of pentaquarks.  However, due to the short life of the resonances, the measure of their magnetic moments may be quite involved.

The magnetic moments of quarks are related to their masses ($\mu_q=\frac{Q_q}{2m_q}$, being $Q_q$ the charge and $m_q$ the mass of the constituent quark) so that they are model-dependent as well and one has to take values consistent with the constituent quark masses that one uses. We will use in this article the values from \cite{Gasiorowicz:1981jz} which are given in Table~\ref{qmoments} and obtained using the masses that fit the low-lying baryon spectrum (Table~\ref{cmasses}).
\begin{table}[h]
\begin{center}
\begin{minipage}{0.45\textwidth}
\begin{center}
\begin{tabular}[t]{|c|c|}\hline
Flavor&Moment\\\hline\hline
u &  1.863\\ \hline
d & -0.931\\ \hline
s & -0.582\\ \hline
\end{tabular}
\caption{\small Magnetic moments  (in $\mu_N$ units) for light quarks (from \cite{Gasiorowicz:1981jz}).}
\label{qmoments}
\vspace{-2.5em}
\end{center}
\end{minipage}
\end{center}
\end{table}

We may calculate the magnetic moments of composite particles with the assumption that these are the vector sum of those of their constituent quarks plus orbital contributions. Since the diquark has spin zero, its spin contribution to the magnetic moment from is nule and the magnetic moment operator for this model is:
\begin{equation}
\vec{\mu}_{Q^2\overline{Q}-Q^2}=\vec{\mu}_{Q^2\overline{Q}}+g_l\vec{l}=g_{Q^2\overline{Q}}\vec{\frac{1}{2}}+g_l\vec{1}
\end{equation}
where $g_{l}=\frac{g_{lQ^2}m_{Q^2\overline{Q}}+g_{lQ^2\overline{Q}}m_{Q^2}}{m_{Q^2}+m_{Q^2\overline{Q}}}$. To evaluate the spin contribution of the triquark we have to consider (\ref{triquark}). The first step is to evaluate the magnetic moments of the quark pairs. The magnetic moment corresponding to $|Q^2\rangle_1$ is nule (the two quarks are coupled to spin zero) and therefore the magnetic moment of the triquark in the configuration $|Q^2\overline{Q}\rangle_1$ reduces the one of the antiquark ($\mu_{|Q^2\overline{Q}\rangle_1}=-\mu_s$ for antiquark $\overline{s}$). The spin of the two quarks in $|Q^2\rangle_2$ are coupled symmetrically and the resulting magnetic moment is $\mu_u+\mu_d$. We couple the spin one of this two quarks to the spin of the antiquark by using the $SU(2)$ CG coefficients to obtain $\mu_{|Q^2\overline{Q}\rangle_2}=\frac{1}{3}(2(\mu_u+\mu_d)-\mu_{\overline{s}})$. Once we know the magnetic moments corresponding to all the functions entering in (\ref{triquark}) it is easy to calculate the moment corresponding to the triquark. For a flavor content $uudds$ it is:
\begin{equation}
\mu_{Q^2\overline{Q}}=(0.582)^2(-\mu_s)+(0.813)^2(\frac{1}{3}(2(\mu_u+\mu_d)+\mu_s))
\end{equation}

The last step is to couple the angular momentum of the relative p-wave ($l=1$) of the triquark-diquark system. For the masses of the triquark and the diquark we apply expressions analogous to (\ref{massformula}) including the color-magnetic interaction term and without orbital contribution. The magnitude of the magnetic moments of pentaquarks depends therefore on how the spin $\frac{1}{2}$ of the antiparticle couples with the relative orbital angular momentum $1$ of the diquark-triquark system, i.e, on the CG coefficients of SU(2). Noting as $|S,S_z\rangle$ the total spin of the pentaquark state, the magnitude of the magnetic moment for the triquark-diquark configuration is:
\begin{equation}
\mu=\langle S,S|\hat{\mu}_{Q^2\overline{Q}} + g_l\hat{l}_{Z}|S,S\rangle
\end{equation}
For the spin $S=\frac{1}{2}$ and $S=\frac{3}{2}$ pentaquark states we obtain respectively:
\begin{equation}
\mu_{\frac{1}{2}}=\frac{1}{3}(2g_l-\mu_{Q^2\overline{Q}}) \hspace{2em}
\mu_{\frac{3}{2}}=g_{lQ^2\overline{Q}}+g_l
\end{equation}
Using the values of Table~\ref{qmoments} for the quark magnetic moments we obtain the magnetic moments for pentaquarks in the triquark-diquark configuration of Table~\ref{tridimom}.
\begin{table}[h]
\begin{center}
\begin{tabular}[t]{|c|c|c|}\hline
\begin{picture}(1 ,14)(0,0)
\scriptsize
\put(-18,-4){$Content$}
\put(20,-1){\line(-3,1){39}}
\put(-1,7){$Spin$}
\end{picture}
&$\rule[-6pt]{0pt}{10pt}$$\frac{1}{2}$&$\frac{3}{2}$\\ \hline\hline
$uudd\overline{s}$&$0.362$&$1.164$\\ \hline
$uuss\overline{d}$&$0.203$&$1.159$\\ \hline
$ddss\overline{u}$&$-0.588$&$-1.897$\\ \hline
\end{tabular}
\end{center}
\vspace{-1em}
\caption{\small Magnetic moments of pentaquarks in the triquark-diquark model (in $\mu_N$ units).\vspace{-1.5em}}
\label{tridimom}
\end{table}

Calculations for the magnetic moments of the spin $\frac{1}{2}$ states using similar quark cluster models with other configurations (in particular for the models of \cite{Jaffe:2003sg}, \cite{Shuryak:2003zi} and \cite{Karliner:2003dt}) may be found in \cite{Liu:2003ab}.

The experimental determination of the parity and the magnetic moments of pentaquarks may be the key for the understanding of their internal structures in constituent quark terms. These measurements are still to be done, but in particular the measure of the magnetic moments is probably quite involved due to the short lifetime of the resonances and is not expected to be realized in the next few years.
\vspace{-1em}

\section*{Charmed pentaquarks}
The triquark-diquark model worked out in this article may be easily extended for charmed pentaquarks of quark content analogous to the considered previously with the replacement $s\rightarrow c$. These states would be labelled $\Theta^0_c$($uudd\overline{s}$), $\Xi^0_c$($ddcc\overline{u}$) and $\Xi^{+++}$($uucc\overline{d}$). It is also possible the formation of charmed pentaquarks containing strange quarks as well, but we will not consider it here.

The masses of charmed quarks in Table \ref{cmasses} have been taken from \cite{Gasiorowicz:1981jz} and are obtained by fitting the charmed baryon spectrum for mesons and baryons respectively.

The color-magnetic interaction term of a quark pair is proportional to the inverse of the masses of the two quarks, so that if one of the quarks is a charm quark this term is negligible compared to the constituent masses and to the color-magnetic interaction between light-quark pairs. Therefore, in the evaluation of the color-magnetic interaction for charmed baryons it suffices to take into account the terms corresponding to light quark pairs. 

The reduced mass of the triquark-diquark system for charmed pentaquarks is quite different from the 410 MeV of the $c\overline{s}$ system from \cite{Karliner:2003dt}. Considering this we obtain an estimation of the rotational energy for charmed pentaquark scaling the estimation for this system ($\Delta E = \frac{\mu_{c\overline{s}}}{\mu}\Delta_{c\overline{s}}$).

The results of applying (\ref{massformula}) to the lowest-lying configurations  for these charmed pentaquarks are given in Table \ref{charmed}.
\begin{table}[h]
\begin{center}
\begin{minipage}{0.45\textwidth}
\begin{center}
\begin{tabular}[t]{|c|c|}\hline
&Mass\\ \hline\hline
$\Theta_c$&3028\\ \hline
$\Xi_c$&4030\\ \hline
\end{tabular}
\caption{\small Estimate (in MeV) of some charmed pentaquark masses using the triquark-diquark model proposed here and a rotational energy from \cite{Karliner:2003dt}.}
\label{charmed}
\vspace{-1.5em}
\end{center}
\end{minipage}
\end{center}
\end{table}
The mass of the $\Theta^0_c$ is quite close to the mass \mbox{(3.17 GeV)} of the system reported in an old experimental paper \cite{Amirzadeh:1979qi}, which was suggested to be a possible pentaquark state. Estimations for charmed pentaquarks using the diquark-diquark-antiquark model of \cite{Jaffe:2003sg} may be found in \cite{Cheung:2003de}.
\vspace{-1em} 

\section*{Conclusions}
Recently several experimental collaborations have reported observations of narrow resonances with the quantum numbers of exotic pentaquark states. The aim of this article was to study the stability of the constituent quark clusters building up pentaquarks by means of the color-magnetic interaction between constituent quarks. The evaluation of the spin-spin color interaction has been taken from bag models \cite{Mulders:1979ea,Aerts:1979hn}. According to this the most stable configuration for pentaquarks is the triquark-diquark model that has been proposed.

The diquark ($\Delta=-2$) has a completely antisymmetric wave function in flavor, spin and color, so that they are quarks of different flavor ($ud$ in the case of the $\Theta^+$) coupled to spin zero and $\overline{3}$ of color.

The physical states of a pentaquark system diagonalize the mass operator (\ref{massformula}), i.e, the group theoretical factor $\Delta$. For the triquark, $\Delta$ is not diagonal in the basis of product states of two-quark clusters times the antiquark state \cite{Mulders:1979ea, deSwart:1981fn}. The eigenvector corresponding to the eigenvalue $\Delta=-5.42$ is a linear combination (\ref{triquark}) of two different two-quark states from (\ref{diquarks}), the one of the diquark ($\Delta=-2$) and the two-quark state of $\Delta=-\frac{1}{3}$ which quarks are coupled antisymmetrically in flavor but symmetrically in spin and color.

Both two-quark systems couple with the antiquark into a spin $S=\frac{1}{2}$ and $3_C$ color representation. Since their wave functions are completely antisymmetric under the interchange of identical particles, the wave function of the triquark is antisymmetric as well. The wave function of the diquark is also antisymmetric, so that to ensure total antisymmetry under the interchange of identical particles triquark and diquark must be in an antisymmetric spatial wave function or relative p-wave ($l=1$). Therefore, the final state is a singlet in color, belongs to the antidecuplet representation of flavor and may have spin-parity state $S^{\small P}=\frac{1}{2}^+$ or $S^{\small P}=\frac{3}{2}^+$.

Using the constituent quark masses that fit the \mbox{s-wave} hadron spectrum and the color-magnetic interaction from bag model calculations we have estimated the masses of pentaquarks (and some charmed pentaquarks) in this configuration with considerable agreement with the observed masses. The model also explains qualitatively the narrow width of the resonances. Finally a prediction for the magnetic moments of pentaquarks has been worked out using this triquark-diquark model.

\setlength{\textheight}{0.7\textheight}

\section*{Acknowledgments}
This article is result of my work as a M. Sc. Student under the supervision of \emph{P. J. Mulders} and \emph{D. Boer}. I would like to thank them for their advices and their constant support.

\thispagestyle{empty}
\end{document}